\documentclass[12pt]{iopart}

\usepackage{graphicx,iopams}

\begin{document}

\title[Effective Hamiltonian parameters of SrCl$_{2}$:Yb$^{2+}$ and CsCaBr$_{3}$:Yb$^{2+}$]{Effective-Hamiltonian parameters for \emph{ab initio}
  energy-level calculations of SrCl$_{2}$:Yb$^{2+}$ and CsCaBr$_{3}$:Yb$^{2+}$}

\author{Alexander J. Salkeld}
\address{Department of Physics and Astronomy, University of
  Canterbury, PB 4800, Christchurch 8140, New Zealand}

\author{Michael F. Reid}
\address{Department of Physics and Astronomy and MacDiarmid
  Institute for Advanced Materials and Nanotechnology, University of
  Canterbury, PB 4800, Christchurch 8140, New Zealand}
\ead{mike.reid@canterbury.ac.nz}

\author{Jon-Paul R. Wells}
\address{Department of Physics and Astronomy, University of
  Canterbury, PB 4800, Christchurch 8140, New Zealand}

\author{Goar S\'{a}nchez-Sanz}
\address{Institute of Organic Chemistry and Biochemistry, Gilead Sciences Research Center \& IOCB, Academy of Sciences, Prague, Czech Republic}

\author{Luis Seijo} 
\address{Departamento de Qu\'{i}mica and 
Instituto Universitario de Ciencia de Materiales Nicol\'{a}s Cabera, 
Universidad Aut\'{o}noma de Madrid,
28049 Madrid, Spain}

\author{Zoila Barandiar\'{a}n}
\address{Departamento de Qu\'{i}mica and 
Instituto Universitario de Ciencia de Materiales Nicol\'{a}s Cabera, 
Universidad Aut\'{o}noma de Madrid,
28049 Madrid, Spain}

\begin{abstract}

  Calculated energy levels from recent \emph{ab initio} studies of the
  electronic structure of SrCl$_{2}$:Yb$^{2+}$ and
  CsCaBr$_{3}$:Yb$^{2+}$ are fitted with a semi-empirical
  ``crystal-field'' Hamiltonian, which acts within the model space
  $4f^{14} + 4f^{13}5d + 4f^{13}6s$.  Parameters are obtained for the
  minima of the potential-energy curves for each energy level and also
  for a range of anion-cation separations. The parameters are compared
  with published parameters fitted to experimental data and to
  atomic calculations. The states with significant $4f^{13}6s$ character
  give a good approximation to the impurity-trapped exciton states that
  appear in the \emph{ab initio} calculations.

\end{abstract}

\submitto{\JPCM}

\section{Introduction}

The energy levels of lanthanide ions in solids, which give rise to transitions from the UV to the IR regions of the
electromagnetic spectrum, have important applications, such as lighting, lasers and
scintillators. The vast majority of energy-level calculations for these
materials make use of a ``crystal-field'' Hamiltonian, though there have
been some \emph{ab initio} calculations \cite{NN86b}. The Hamiltonian
was originally developed for the $4f^{N}$ configuration of lanthanide
ions \cite{CGRR89,LiJa05}. It has also been applied to actinides and
extended to include the $4f^{N-1}5d$ configuration of both divalent
\cite{PiBrMc67,PaNiChTa06,KaUrRe07,PaDuTa08}, and trivalent
\cite{PiReWeSoMe02,PiReBuMe02,BuRe07} lanthanide ions.  Spectra
involving the $4f^{N-1}5d$ configuration are generally in the UV and VUV
region. Due to changes in bonding, and hence bond length, between
$4f^{N}$ and $4f^{N-1}5d$, transitions between configurations generally
involve broad vibronic bands, though in materials with heavy ligands it
is possible to observe vibronic progressions, particularly for divalent
ions \cite{PiBrMc67,PaNiChTa06,KaUrRe07,PaDuTa08}.

In addition to the $4f^{N-1}5d$ configuration, other excited
configurations play a role in high-energy spectra.  The $4f^{N-1}6s$
configuration generally overlaps the $4f^{N-1}5d$ configuration
\cite{PaDuTa08,MaBrRyWiSwHeGu12}, and there are also states involving
charge transfer between the lanthanide and the other ions in the
material. Charge can be transferred from a ligand to the lanthanide,
giving broad charge-transfer absorption bands \cite{PiReWeSoMe02}. On
the other hand, charge from the lanthanide may become delocalized,
leading to an impurity trapped exciton state (ITE). The delocalization
gives a very large change in bond length relative to the $4f^{N}$
configuration and so these states may sometimes be directly observed by
their broad, red-shifted, ``anomalous'' emission bands
\cite{MoCoPe89,MoCoPe91,McPe85a,McPe85b,Do03a}, and are also thought to
play a crucial role in non-radiative relaxation from states in the
$4f^{N-1}5d$ configuration \cite{GrMa08,MaGrCaBeBo09}.

There is little direct knowledge of the electronic and geometrical
structure of ITEs in lanthanide materials. We have recently reported the
use of two-colour (UV - IR) excitation to probe the energy-level
structure of ITEs in the CaF$_2$:Yb$^{2+}$ system
\cite{ReSeWeBeMeSaDuRe11}. The energy levels of the ITE were
modelled using a simple crystal-field model where the localized
$4f^{13}$ electrons were coupled to an $s$ electron, since the
delocalized electron would be expected to be hybridized Yb$^{2+}$ $6s$
and Ca$^{2+}$ $4s$ orbitals.

\begin{figure}
\includegraphics[width=0.9\columnwidth]{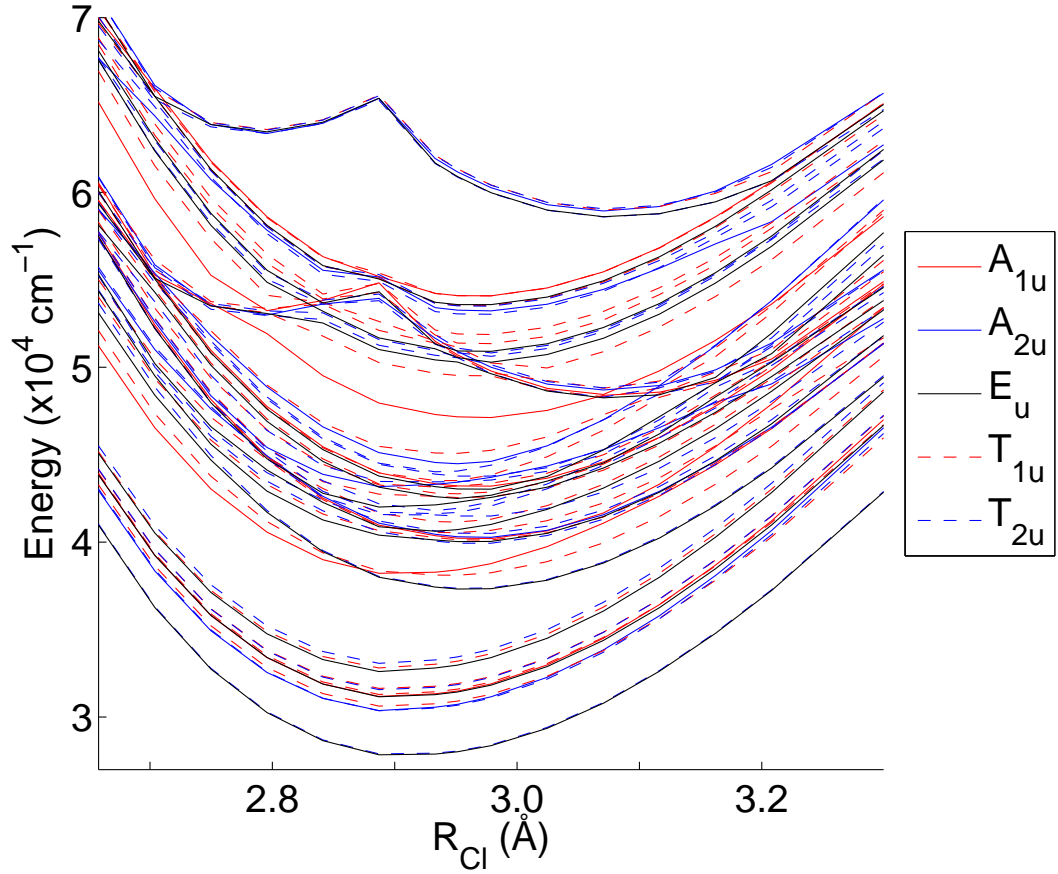}
\caption{\label{fig:EPlot} (Color online.)  Calculated energy levels as
  a function of anion-cation separation ($R_{Cl}$) for the SrCl$_{2}$:Yb$^{2+}$ system
  of S\'anchez-Sanz et al.~\cite{SaSeBa10a}. The double-well potential curves occur on
  states with predominantly $6s$ character.
  The $4f^{14}$ curve, which has a minimum at 2.954~\AA, is omitted.} 
\end{figure}

ITE states have also been a subject of \emph{ab initio} calculations in
various materials \cite{OrSeBa07,SaSeBa10a,SaSeBa10b}. Relevant to this paper, these calculations determine potential energy curves of the electronic states of the materials involved. The most comprehensive calculation is for the SrCl$_2$:Yb$^{2+}$ system
\cite{SaSeBa10a,SaSeBa10b}. This material does not exhibit excitonic
emission \cite{PaDuTa08}, but the absorption bands vary considerably in
width \cite{PiBrMc67}, indicating that the excited states do not all
have the same equilibrium bond length. In the calculations of S\'anchez-Sanz et al.~\cite{SaSeBa10a} there is a double-well energy curve for some
states, as shown in Fig.~\ref{fig:EPlot}. At long anion-cation
separations these states
have predominantly $6s$ character, but at shorter distances they become
$a_{1g}$ symmetry combinations composed of $5s$ orbitals on the
next-nearest-neighbor Sr$^{2+}$ ions, with a contribution from interstitial charge
density. The combined $4f^{13}a_{1g}$ configuration has $A_{1u}$ symmetry. It was demonstrated by S\'anchez-Sanz et al.~\cite{SaSeBa10b} that all of the states were
crucial in explaining the different band-widths observed in the
absorption spectrum \cite{PiBrMc67}.  In this calculation the $A_{1u}$
potential curve minima have higher energies than the minima of the $6s$
or $5d$ states, but could be described as a ``precursor'' to the
exciton states that occurs in
SrF$_{2}$:Yb$^{2+}$ and CaF$_{2}$:Yb$^{2+}$ at lower energy than the
$4f^{13}5d$ configuration \cite{MoCoPe89,Do03a}.

The calculations of S\'anchez-Sanz et al.~\cite{SaSeBa10a} were directly compared with
absorption \cite{PiBrMc67} and emission \cite{PaDuTa08} spectra. While
this gives a good match to the data, it does not provide information directly about the magnitudes of the physical interactions involved, and hence this calculation cannot be compared to the crystal-field calculation of Pan et al.~\cite{PaDuTa08}. 
The advantage of determining the crystal-field parameters of the system is that the corresponding interactions have predictable behavior across the lanthanide series, allowing for the parameters determined for this system to be extrapolated to other lanthanide systems. 
In simple systems with high symmetry and one valence electron, such as the
BaF$_2$:Ce$^{3+}$ system \cite{PaScBaSe06} it is relatively easy to relate the
energy levels to crystal-field parameters. However, for a many-electron
system such as SrCl$_{2}$:Yb$^{2+}$ the relationship is much more
complicated because there are many more parameters in the effective Hamiltonian.

Recent work has shown that it is possible to construct an effective
Hamiltonian matrix from \emph{ab initio} calculations, and
hence determine crystal-field and other parameters by a straightforward
projection technique~\cite{ReDuZh09,ReHuFrDuXiYi10,HuReDuXiYi11}. There the relevant parameters were extracted directly from the \emph{ab initio} Hamiltonian. However, those works were on systems with a single
valence electron. In systems with more than one valence electron the
labelling of the states in terms of the states used in the effective
Hamiltonian approach is not straightforward. Therefore, in this work we
use the approach of fitting the effective Hamiltonian parameters to
calculated energy levels, as in Duan et al.~\cite{DuReXi07}.

The focus of this paper is to describe the behavior of exciton energy
levels using the eigenvalues of a semi-empirical effective Hamiltonian
operator acting on a model space of the SrCl$_{2}$:Yb$^{2+}$ system. The
energy level structure calculated by S\'anchez-Sanz et al.~\cite{SaSeBa10a} is used as
model data to test whether the chosen effective Hamiltonian operator can
reproduce the ``precursor'' exciton results, and to find values for the
physical parameters describing the effective Hamiltonian in this case. \emph{Ab initio}
calculations of a second crystal, CsCaBr$_{3}$:Yb$^{2+}$, are also examined.
The crystal CsCaBr$_{3}$:Yb$^{2+}$ is not excitonic and does not exhibit precursor exciton behavior, 
so it is of interest to study the similarities and differences between the excitonic and non-excitonic cases. The motivations to determine the parameters for these systems are two-fold; as a supplement they reinforce the \emph{ab initio} calculations, allowing comparison to the expected magnitudes of electron interactions. Secondly, the parameters should be comparable in other lanthanide ions, and the ``free ion'' parameters transferable between materials. Therefore, the analysis will be of interest beyond the systems considered here.

\section{Effective Hamiltonian}

For a Hamiltonian $H$, with eigenstates $\psi_i$ and eigenvalues $E_i$, 
an effective Hamiltonian $H_\mathrm{eff}$ is defined so that within a
subspace of the full Hamiltonian it has eigenstates $\phi_i$ with the
same eigenvalues as $H$: 
\begin{eqnarray}
H \psi_i &=& E_i \psi_i , \\
H_\mathrm{eff} \phi_i &=& E_i \phi_i .
\end{eqnarray}
General discussions of effective Hamiltonians may be found in the
literature \cite{Bra67,lindgren,HF93}. In many cases the effective
Hamiltonian is expressed in terms of parameters that are understood to
represent various physical interactions. This is the case for a ``crystal-field''
Hamiltonian applied to lanthanide ions where operators represent
atomic interactions such as the Coulomb and spin-orbit interactions, and
also ``crystal-field'' interactions with the surroundings.  The analysis
of lanthanide energy levels in solids has, since the 1960's, made use of
such a ``crystal-field'' effective Hamiltonian (e.g.~\cite{CGRR89}) for the
$4f^N$ configuration.  This was extended to the $4f^{N-1}5d$
configuration by various workers \cite{PiBrMc67,PiReWeSoMe02,PiReBuMe02}
and a general review is given by Burdick and Reid~\cite{BuRe07}. 
%

Yb$^{2+}$ has a ground configuration $4f^{14}$. This provides only one
energy level, a $^1S_0$ state with energy $E_{\mathrm{avg}}(f)$.

In the excited configuration $4f^{13}5d$ two- and three-body
interactions between the $4f$ electrons do not contribute, and the
Hamiltonian for these states is
\begin{eqnarray}
\label{effHamFD}
H_{4f^{13}5d}& = & E_{\mathrm{avg}}(f) + \Delta_E(fd) \nonumber \\
& +&\zeta(f)A_{\mathrm{so}}(f) \nonumber
 +\sum_{k=4,6}B^k_q{(f)}C^{(k)}_q{(f)} \nonumber \\
& +&\zeta(d)A_{{so}}(d)+\sum_{k=2,4}F^k{(fd)}f_k{(fd)} \nonumber\\
& +&\sum_{j=1,3,5}G^k{(fd)}g_k{(fd)}+\sum_{k=4}B^k_q{(d)}C^{(k)}_q{(d)}.  
\end{eqnarray}
Here $\Delta_E(fd)$ is the average energy of the $4f^{13}5d$
configuration relative to $4f^{14}$. The $\zeta(f)$ and $\zeta(d)$
parameters, and corresponding $A_{so}$ operators, comprise the
spin-orbit effects on the $4f$ electrons and $5d$ electron
respectively. The $F^{k}(fd)$ and $G^{k}(fd)$ parameters are the direct and exchange Slater parameters
for the Coulomb interaction between electrons in different
shells. $B^{k}_{q}$ parameters describe the crystal field effects for
the appropriate electrons.

Impurity sites in SrCl$_{2}$:Yb$^{2+}$ and CsCaBr$_{3}$:Yb$^{2+}$ have eight- and six-coordinate cubic symmetry respectively. Cubic symmetry at the impurity site reduces the number of crystal field parameters to two non-zero parameters for the 4f crystal field, and to one non-zero 5d crystal field parameter \cite{BuRe07,ReNe00}. Due to the different coordinations, the values of these parameters should differ by a sign between SrCl$_{2}$:Yb$^{2+}$ and CsCaBr$_{3}$:Yb$^{2+}$.

In order to extend the model space to include states involving $6s$ orbitals of the Yb$^{2+}$ ion, the $4f^{13}6s$ effective Hamiltonian,
%
%
\begin{eqnarray}
\label{effHam_a}
H_{4f^{13}6s}&=& E_{\mathrm{avg}}(f) + \Delta_E(fs) + \zeta(f)A_{\mathrm{so}}(f)\nonumber\\
&+&\sum_{k=4,6}B^k_q{(f)}C^{(k)}_q{(f)} + G^3{(fs)}g_3{(fs)},
\end{eqnarray}
is added to the $4f^{13}5d$ effective Hamiltonian. Here the $\Delta_E(fs)$ term is the average energy of the $4f^{13}6s$
configuration relative to $4f^{14}$. The $g_3(fs)$ parameter is the only non-zero Coulomb
interaction term acting between $4f$ and $6s$ electrons. Duplicate parameters $\zeta(f)$ and $B^k_q{(f)}$ are set equal to their counterparts in~(\ref{effHamFD}). Finally, an operator accounting for the interaction between $5d$ and $6s$ orbitals is added:
\begin{equation}
\label{effHam_b}
H_{ds}= \sum_{k=2,3}R^k{(ds)}r_k{(ds)},
\end{equation}
where the $R^k$ parameters measure the mixing of states that occurs due to coulomb interactions.


\section{Calculations}

This section is divided into two parts. Firstly, the parameters of the
effective Hamiltonian operator, (Eq.~\ref{effHamFD}--\ref{effHam_b}), are fitted to the 
energy eigenvalues at the minima of the potential curves of S\'anchez-Sanz et al.~\cite{SaSeBa10a, SaSeBa09}. These are the predicted positions of the zero-phonon
lines, so, this calculation is directly comparable to the crystal-field
calculation of Pan et al.~\cite{PaDuTa08} (for SrCl$_{2}$).  Secondly, the same effective Hamiltonian
operator is fitted to the energy eigenvalues calculated at fixed anion-cation separation. This allows for the investigation of the variation in parameter values
with anion-cation separation.

\subsection{Fit to potential curve minima}
\label{minima}

Initial values for the effective Hamiltonian parameters were chosen from
values determined by Pan et al.~\cite{PaDuTa08} for the $4f$ and $5d$
parameters. Values for the $G^3(fs)$, $R^2(ds)$ and $R^3(ds)$ parameters
were calculated from their integral definitions \cite{cowan,lindgren},
using Hartree-Fock wavefunctions as estimates of electron radial
distribution \cite{lindgren}.

\begin{table}
\center
\caption{Parameter values for the effective Hamiltonian of SrCl$_2$:Yb$^{2+}$ 
fitted to experimental observations~\cite{PaDuTa08}; 
calculated for a free Yb$^{2+}$ ion~\cite{cowan}; 
and fitted to ab-initio calculations of S\'anchez-Sanz et al.~\cite{SaSeBa10a}. 
The labels CASSCF, MS-CASPT2 and SO-CI refer to the level of the reference calculation. At CASSCF level, only basic interactions are considered. MS-CASPT2 includes dynamic correlation of electrons, and SO-CI includes spin-orbit interactions. 
All parameter values and uncertainties ($\sigma$) are in cm$^{-1}$. The
standard deviation for the SO-CI SrCl$_2$ 
fit is $\sigma \approx 190$ cm$^{-1}$. 
\label{MinParams}}

\bigskip 

\tiny

\begin{tabular}{@{}lrrrrrrrrrrrrrr}
\br
& \multicolumn{4}{c}{CASSCF} & \multicolumn{4}{c}{CASPT2} & \multicolumn{4}{c}{SO-CI} & Expt.\cite{PaDuTa08} & Atomic \cite{cowan} \\
Parameter & \multicolumn{2}{c}{CsCaBr$_{3}$} & \multicolumn{2}{c}{SrCl$_{2}$} & \multicolumn{2}{c}{CsCaBr$_{3}$} & \multicolumn{2}{c}{SrCl$_{2}$} & \multicolumn{2}{c}{CsCaBr$_{3}$} & \multicolumn{2}{c}{SrCl$_{2}$} &  & \\
& value &$\sigma$ & value &$\sigma$ & value &$\sigma$ & value &$\sigma$ & value &$\sigma$ & value &$\sigma$ \\
\mr
$\Delta_E(fd)$         & $     13905$ & $       275$ & $     12099$ & $       202$ & $     39847$ & $       390$ & $     41873$ & $       351$ & $     39793$ & $       365$ & $     41802$ & $        42$ & $     38382$ & $$ \\
$\zeta(f)$       & $$ & $$ & $$ & $$ & $$ & $$ & $$ & $$ & $      2932$ & $       118$ & $      2939$ & $        43$ & $      2950$ & $      2899$ \\
$\zeta(d)$       & $$ & $$ & $$ & $$ & $$ & $$ & $$ & $$ & $       756$ & $       604$ & $      1166$ & $        43$ & $      1211$ & $      1290$ \\
$F^2(fd)$         & $     21390$ & $       703$ & $     23665$ & $       487$ & $     15946$ & $      1013$ & $     17815$ & $       880$ & $     16795$ & $      2841$ & $     18393$ & $       169$ & $     14355$ & $     23210$ \\
$F^4(fd)$         & $      9837$ & $      1287$ & $     10649$ & $       922$ & $     11403$ & $      1871$ & $     13475$ & $      1700$ & $      9489$ & $      8174$ & $     13099$ & $       449$ & $      7222$ & $     10646$ \\
$G^1(fd)$         & $      8761$ & $       221$ & $      9202$ & $       139$ & $      2916$ & $       437$ & $      4864$ & $       358$ & $      4331$ & $      2412$ & $      5408$ & $        93$ & $      4693$ & $     10059$ \\
$G^3(fd)$         & $      7155$ & $       701$ & $      9062$ & $       560$ & $      7436$ & $      1022$ & $      9403$ & $       946$ & $      6947$ & $      6126$ & $      8901$ & $       454$ & $      5382$ & $      8046$ \\
$G^5(fd)$         & $      6199$ & $       960$ & $      6351$ & $       653$ & $      5655$ & $      1518$ & $      5459$ & $      1251$ & $      7511$ & $      9474$ & $      7165$ & $       544$ & $      4349$ & $      6085$ \\
$\Delta_E(fs)$       & $     21280$ & $       159$ & $      9221$ & $       119$ & $     21357$ & $       228$ & $     11130$ & $       204$ & $     21605$ & $       562$ & $     11093$ & $        45$ & $$ & $$ \\
$G^3(fs)$         & $      2733$ & $       882$ & $      2984$ & $       675$ & $      1498$ & $      1270$ & $      2484$ & $      1138$ & $      2543$ & $      4324$ & $      2604$ & $       505$ & $$ & $      3168$ \\
$R^2(ds)$         & $       922$ & $      2462$ & $      3474$ & $       898$ & $      7934$ & $      2406$ & $      1038$ & $      2066$ & $      -902$ & $      3574$ & $      1990$ & $      1238$ & $$ & $      -305$ \\
$R^3(ds)$         & $     -1030$ & $      4839$ & $       699$ & $      1236$ & $      2655$ & $      2999$ & $      3634$ & $      3379$ & $      8839$ & $      6185$ & $      2449$ & $       900$ & $$ & $      1468$ \\
$B^4(f)^a$         & $       595$ & $       334$ & $       533$ & $       256$ & $      1721$ & $       476$ & $       473$ & $       326$ & $      1381$ & $      1702$ & $      -194$ & $        24$ & $      -725$ & $$ \\
$B^6(f)^b$         & $        89$ & $       173$ & $        41$ & $       123$ & $      -108$ & $       258$ & $       -73$ & $       422$ & $      -187$ & $      1113$ & $      -592$ & $        28$ & $       292$ & $$ \\
$B^4(d)^a$         & $     35199$ & $       245$ & $    -18966$ & $       199$ & $     39763$ & $       328$ & $    -20221$ & $       232$ & $     39639$ & $       853$ & $    -20100$ & $        79$ & $    -20442$ & $$ \\
\br
\end{tabular}

$^a$ {$B^4_0=B^4$,  $B^4_{\pm4}=\sqrt{\frac{5}{14}}B^4$.} \hfill{\ } \\  
$^b$ {$B^6_0=B^6$,  $B^6_{\pm4}=-\sqrt{\frac{7}{2}}B^6$.} \hfill{\ }

\end{table}

Non-linear least-squares regression was used to optimize the
parameters by fitting the energy eigenvalues of the effective
Hamiltonian to the minima of the potential energy curves calculated by S\'anchez-Sanz et al.~\cite{SaSeBa10a, SaSeBa09}. A fit was performed for each level of the calculations presented in these references. For the states with
double-well potential curves, the longer anion-cation separation ($6s$ character) minima were used. The ratios of direct and
exchange Coulomb parameter values were not constrained in any way.

The standard deviation of the fits are calculated via
\begin{equation}
\label{stdDev}
\sigma = \sqrt{\frac{\sum^{N_{pts}}_{i}\left(E_i-x_i\right)^2}{N_{pts}-N_{vars}}},
\end{equation}
where $N_{pts}$ is the number of energies fitted, and $N_{vars}$ is the number of free parameters. 

Additionally, for the spin-orbit inclusive calculation of SrCl$_{2}$:Yb$^{2+}$, (labelled SO-CI), 
the convergence of the fit was tested. To test the convergence of the fit, we took a large number of starting positions in the 15-dimensional parameter space to provide
varying initial values. A number of solutions converging to
local minima in the parameter space were identified. Several local
minima occur around the best standard deviation, $\sigma \approx
190~\mathrm{cm}^{-1}$, varying mostly in $R^2(ds)$ and $R^3(ds)$
parameter values. This is similar to the fit to
SrCl$_2$:Yb$^{2+}$ spectra \cite{PaDuTa08} where $\sigma \approx 174$
cm$^{-1}$ (fitted to $T_{1u}$ states). The averaged parameter values from these solutions are
presented in Table~\ref{MinParams}. 

\subsection{Fit by ligand separation}
\label{lengths1}

\begin{figure*}
\begin{minipage}[t]{0.5\textwidth}
(a)\includegraphics[width=0.9\columnwidth]{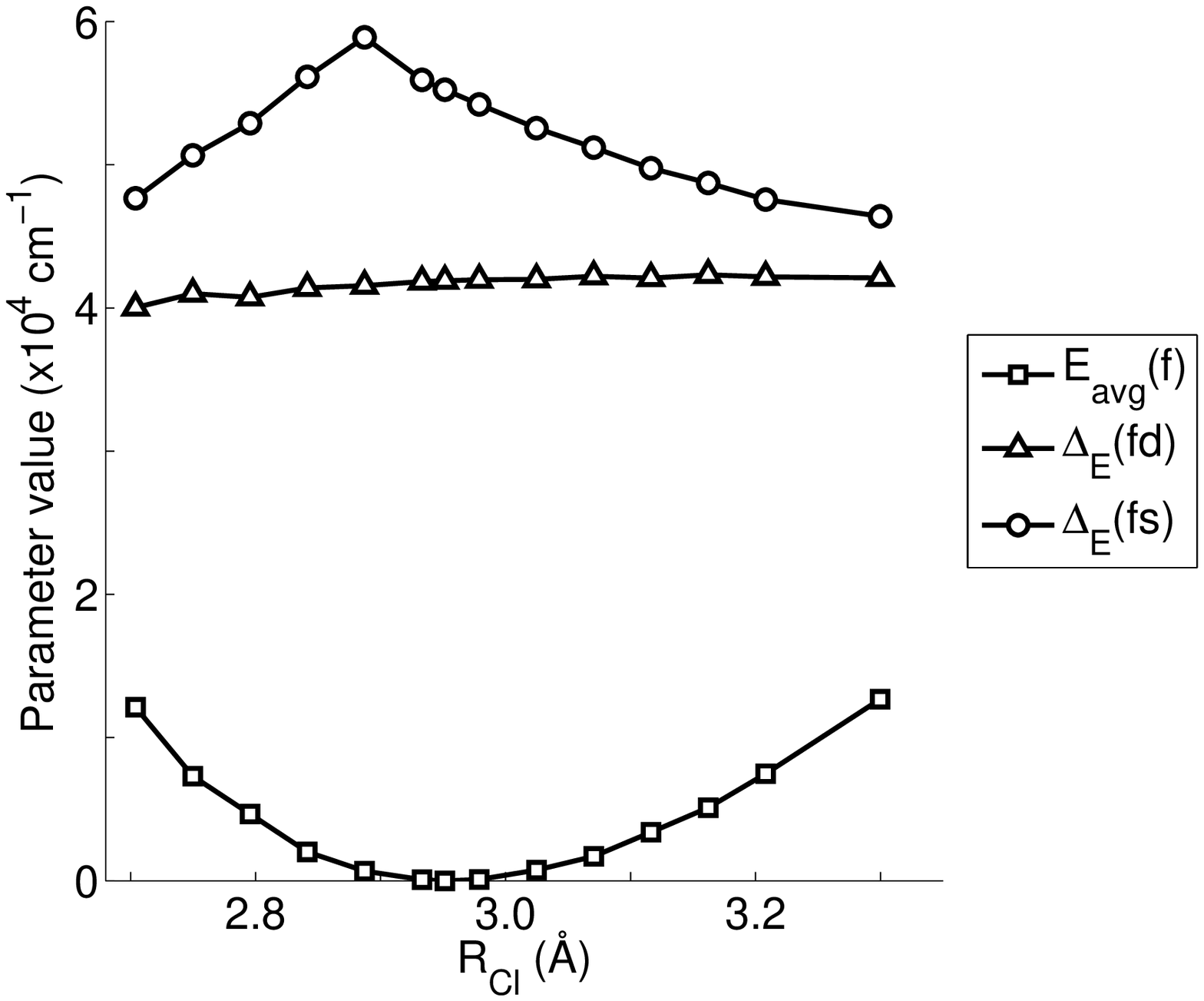}
(b)\includegraphics[width=0.9\columnwidth]{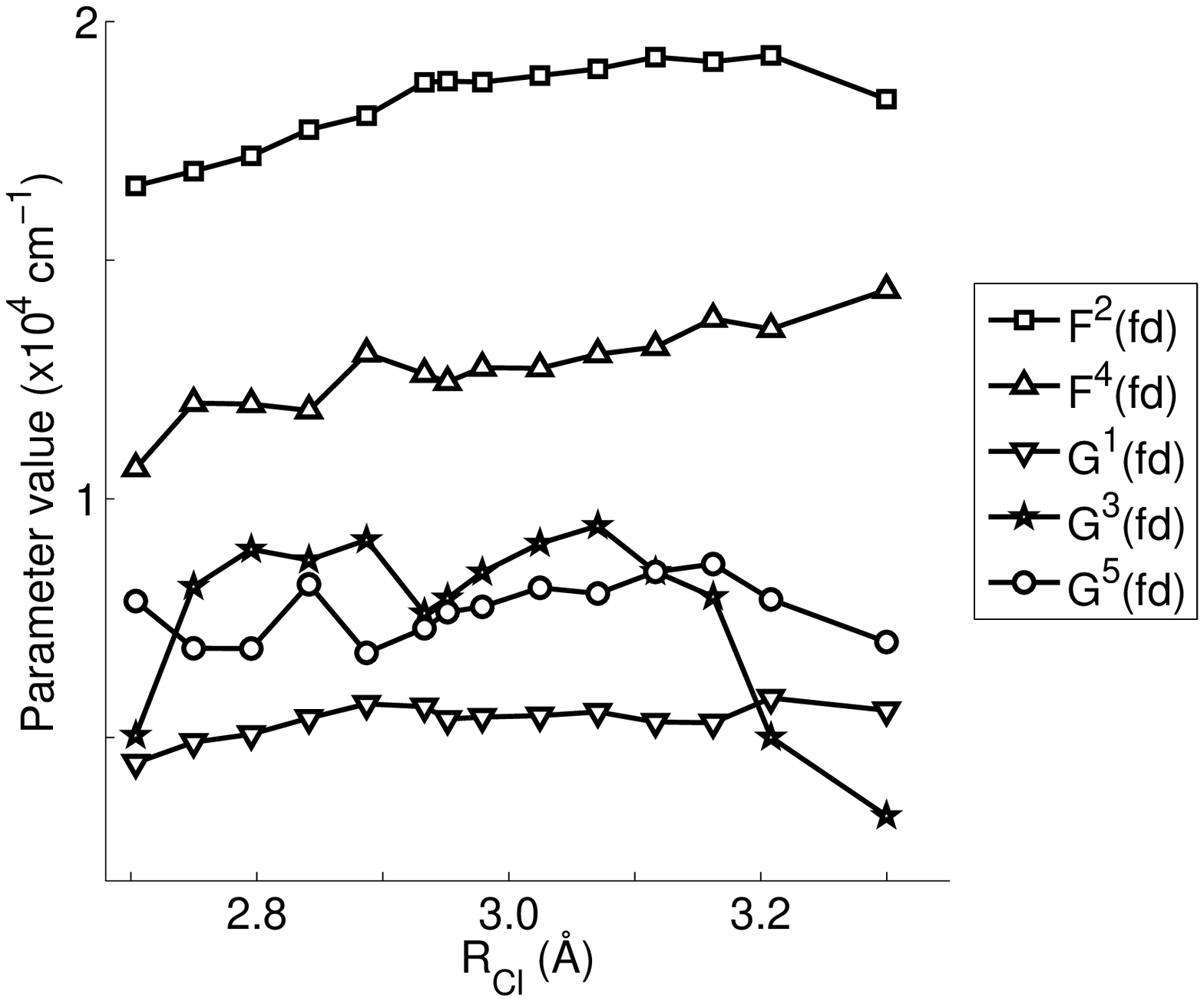}\\
(c)\includegraphics[width=0.9\columnwidth]{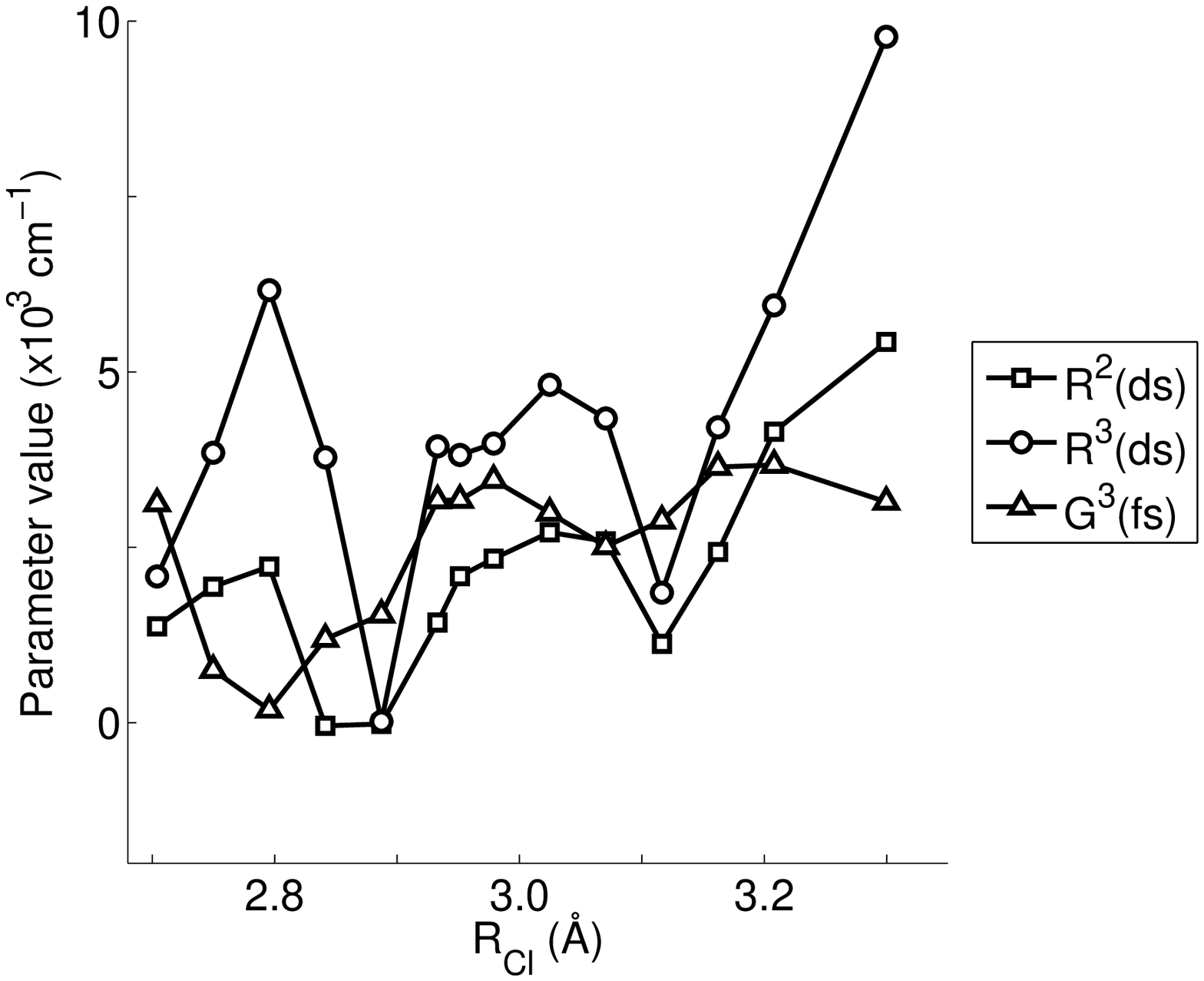}
\end{minipage}
\begin{minipage}[t]{0.5\textwidth}
(d)\includegraphics[width=0.9\columnwidth]{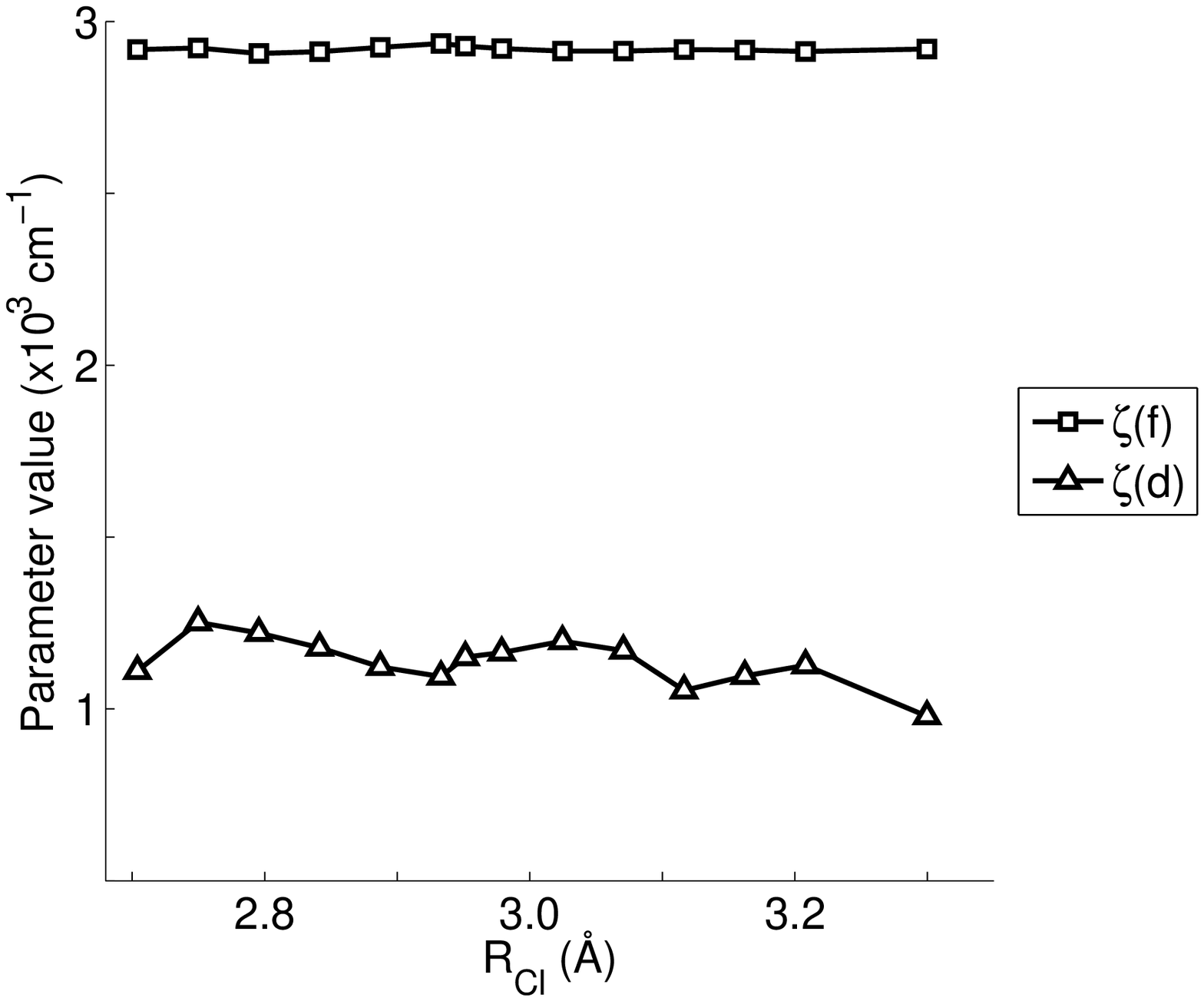}\\  
(e)\includegraphics[width=0.9\columnwidth]{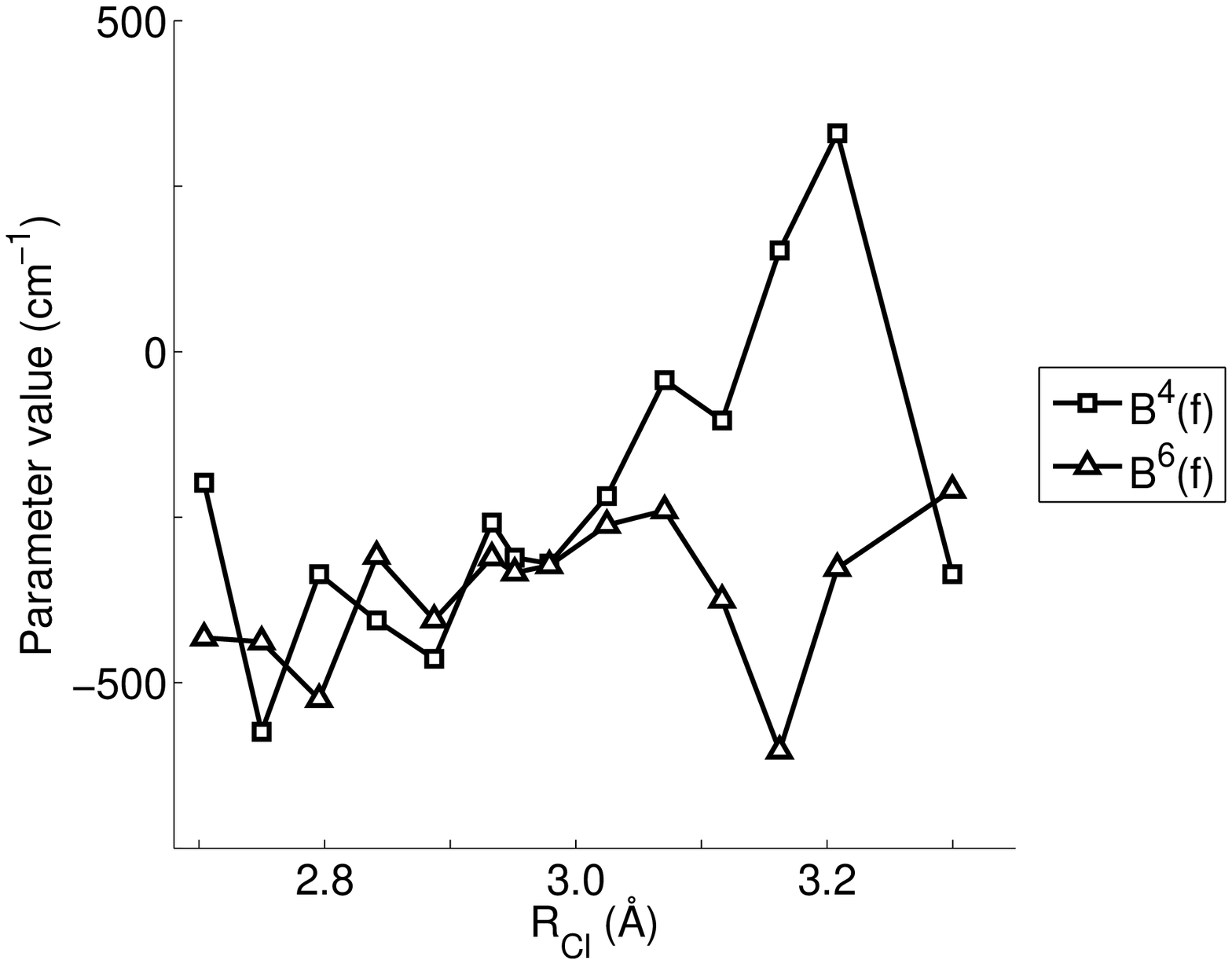}
(f)\includegraphics[width=0.9\columnwidth]{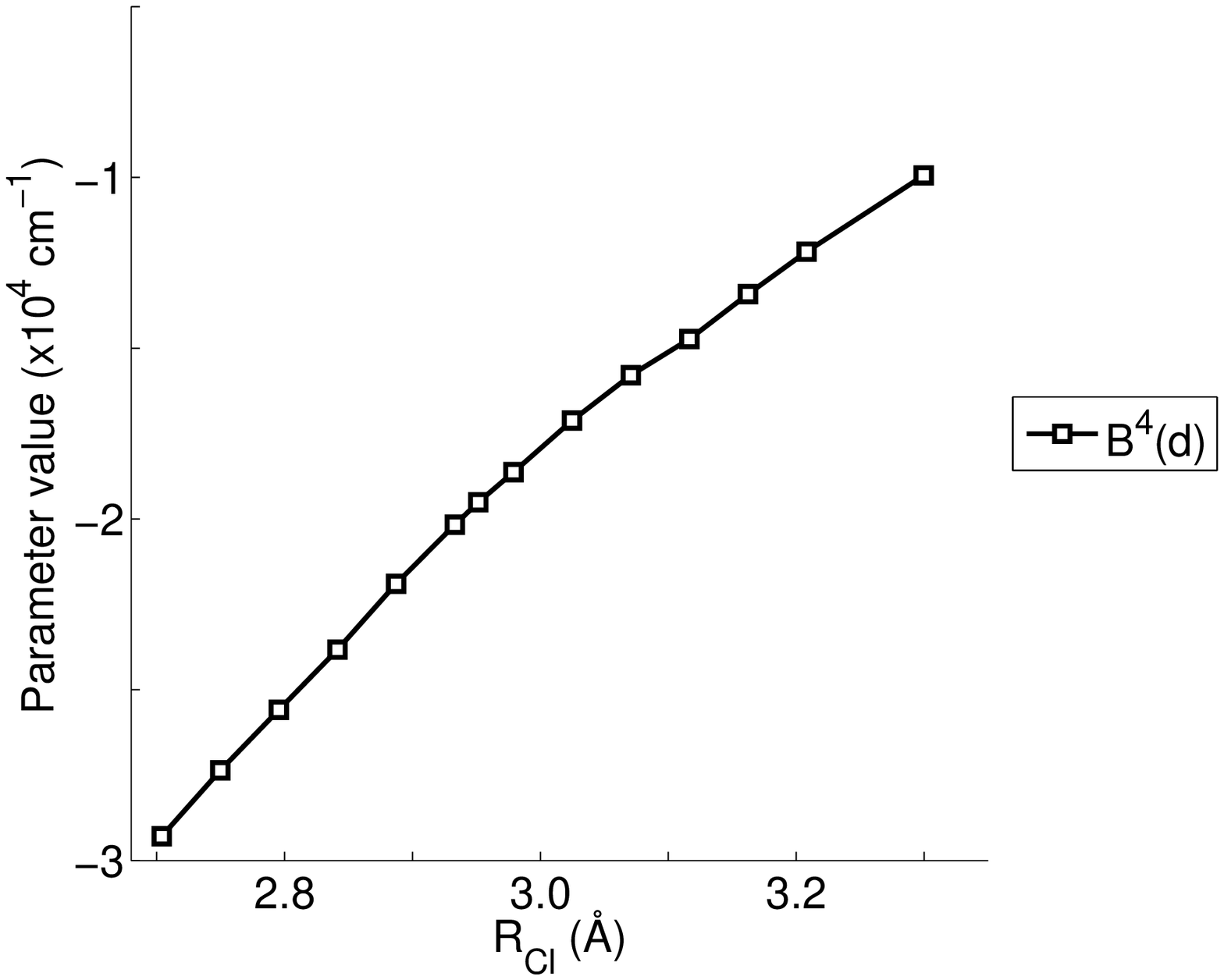}
\end{minipage}
\caption{\label{fig:H1Param} Parameters for the 
  $4f^{14}+4f^{13}5d+4f^{13}6s$ effective Hamiltonian,
  optimized by fitting  to energies of
  SrCl$_{2}$:Yb$^{2+}$ using the spin-orbit level calculation (SO-CI) of S\'anchez-Sanz et al.~\cite{SaSeBa10a}. The plots are split into groups of comparable
  parameters: (a) Average configuration energy parameters. (b) $fd$
  Coulomb parameters.  (c) $fs$ Coulomb parameters.  (d) Spin-orbit
  parameters.  (e) $4f$ crystal-field parameters.  (f) $5d$ crystal-field
  parameter.
}  
\end{figure*}

The effective Hamiltonian operator was also used to fit to the
energies at a fixed anion-cation separation for each 
separation calculated by S\'anchez-Sanz et al.~\cite{SaSeBa10a, SaSeBa09}. The separation 
of 2.9514~\AA\ was chosen first, as the minima of most states occur around this ion separation. 
Thus, the best-fit parameters in Table~\ref{MinParams} could be used as suitable initial parameter
values. The best fit for each anion-cation separation was
subsequently determined using the best fit parameters for an adjacent
separation as initial parameter estimates.

The fitted parameters are plotted against ligand separation for the SrCl$_{2}$:Yb$^{2+}$ SO-CI fits in Fig.~\ref{fig:H1Param}.
Several clear trends are visible in the parameters as they vary with length, holding
even at the point of discontinuity between the $6s$-electron and $a_{1g}$
ITE regimes. These trends are also visible in the fits to other calculation levels, and for the calculations of CsCaBr$_{3}$:Yb$^{2+}$ (not shown).

For SrCl$_{2}$:Yb$^{2+}$, the best fits were obtained at ion separations that correspond to either
minimum of the double-well potential curves. For the longer 
ion separation potential well, the fits ranged in standard deviation from
130~cm$^{-1}$ to 190~cm$^{-1}$, with the best parameter fit to the
energy levels occurring at a separation of 3.1164~\AA. The
shorter separation well, corresponding to $a_{1g}$ exciton-like behavior,
has a minimum standard deviation of $\sigma=156~\mathrm{cm}^{-1}$ at
a separation of 2.7498~\AA. However, the accuracy of the fit decreases
rapidly as the ion separation decreases, with the standard
deviation climbing to $\sigma>500~\mathrm{cm}^{-1}$. Fits for
anion-cation separations shorter than 2.7~\AA\ have not been shown on the plots,
since the parameter values fluctuated wildly. At the point of
discontinuity between the $6s$-electron and $a_{1g}$ exciton regimes
(separation of 2.8414~\AA) there is a reasonable fit to the energies, with a standard deviation of $\sigma=190~\mathrm{cm}^{-1}$.

Similarly, the fixed ion separation fits were performed for the calculations of CsCaBr$_{3}$:Yb$^{2+}$~\cite{SaSeBa09}. These had fewer data points, which had a more consistent accuracy of fit, with a standard deviation of 140 cm$^{-1}$ to 200 cm$^{-1}$. 

\section{Discussion}

For all calculations, the $4f^{13}5d$ parameters converge to a small range of values for each parameter. The largest
variation arises in the $R^2(ds)$ and $R^3(ds)$ parameters. The
matrix representations of the $r_2$ and $r_3$ operators have no diagonal
elements, hence there is only a weak mixing of some $5d$ and $6s$
states. The energies are insensitive to variations in $R^2(ds)$ and $R^3(ds)$ parameters, hence the fit uncertainties of these are
correspondingly large, as shown in Table~\ref{MinParams}. In the SrCl$_{2}$:Yb$^{2+}$ system, the $R^2(ds)$
and $R^3(ds)$ parameters tend to converge to positive and negative
values, of similar magnitude, with equal frequency. These values are more stable for the CsCaBr$_{3}$:Yb$^{2+}$ fits.

\subsection{Fit to potential curve minima}


In Table~\ref{MinParams} the parameters from experimental fits
\cite{PaDuTa08}, atomic calculations \cite{cowan}, and fits to the
\emph{ab initio} potential minima \cite{SaSeBa10a} are shown. The
experimental fit is to a subset of the energy levels, since only
$T_{1u}$ states are accessible by absorption from the ground state
\cite{PiBrMc67}.  However, experimental fits to the SrF$_2$:Sm$^{2+}$
spectrum \cite{KaUrRe07} and the SrF$_2$:Eu$^{2+}$ spectrum
\cite{PaNiChTa06} are also available, and these are in broad agreement
with the parameters for SrF$_2$:Yb$^{2+}$ of Pan et al.~\cite{PaDuTa08}.

We begin by considering the parameters relevant to the $4f^{13}5d$
configuration.  The experimental, atomic, and \emph{ab initio}
spin-orbit parameters $\zeta(f)$ and $\zeta(d)$ of SrCl$_{2}$ and $\zeta(f)$ of CsCaBr$_{3}$ are very
similar. The atomic calculations and all of the \emph{ab initio} calculations overestimate
the $fd$ Coulomb parameters $F^k(fd)$ and $G^k(fd)$. This is consistent
with many studies of trivalent lanthanide ions \cite{CGRR89,LiJa05}, and is an indication that electron-correlation
effects are not fully accounted for by the \emph{ab initio}
calculations. Since the atomic calculations are at the Hartree-Fock
level, it is expected that they would overestimate these parameters
\cite{MoRa71,DuReXi07}.

There is excellent agreement between \emph{ab initio} and experimental
values for the $5d$ crystal-field parameter $B^4(d)$. However, the $4f$
crystal-field parameters show poor agreement, with the $B^6(f)$ parameter
disagreeing in sign.  This may be simply a problem of insensitivity, and
it is notable that in the experimental fits to SrF$_2$:Eu$^{2+}$ the
$B^6(f)$ parameter was also negative. A key aspect is that
the $4f$ crystal-field parameters are up to two orders of magnitude weaker
than the other electronic interactions. Thus, the individual
contributions from these crystal-field operators are easily lost in the
noise of the fit to the other parameters. 


Now we turn to the $4f^{13}6s$ configuration of SrCl$_{2}$:Yb$^{2+}$. As discussed above, the
$R^2(ds)$ and $R^3(ds)$ parameters are not well-determined. However, the
$G^3(fs)$ parameter is well-determined, and the \emph{ab initio} value is similar to the
atomic value. This is to be expected since at the minima fitted the
\emph{ab initio} calculation predicts predominantly $6s$ character for
the excited electron \cite{SaSeBa10a}.

\subsection{Fit by ligand separation}

Parameters as a function of anion-cation separation for the SrCl$_{2}$:Yb$^{2+}$ SO-CI calculation are shown in
Fig.~\ref{fig:H1Param}. Similar analyses were performed on all levels of the calculations presented by S\'anchez-Sanz et al.~\cite{SaSeBa10a, SaSeBa09}. The $E_{\mathrm{avg}}(f)$ parameter reflects the
potential well for the system. The $4f^{13}6d$ configuration average
follows almost the same curve, so $\Delta_E(fd)$ is approximately
constant.  The $4f^{13}6s$ average has a discontinuity at 2.85~\AA, as
seen in $\Delta_E(fs)$, reflecting an avoided crossing between localized
and delocalized states of the excited electron. At longer distances
these states are predominantly $4f^{13}6s$, but at shorter distances they
have significant delocalization, and can be considered an ITE state \cite{SaSeBa10a}.

The $4f^{13}6d$ spin-orbit parameters show little variation with anion-cation separation. However, the $F^{k}$(fd) and $G^{k}$(fd) parameters show a
general decline with decreasing ion separation which is consistent with a
nephalauxetic-effect interpretation, where increased bonding
delocalizes the electrons, and reduces the Coulomb interactions \cite{ScJo58}.
This pattern is observed in the Coulomb parameters for all calculation levels of both crystals.

The SrCl$_{2}$ $4f$ crystal-field parameters $B^4(f)$ and $B^6(f)$ show wide
fluctuations. As noted above, they appear to be poorly
determined. However, between 2.8~\AA\ and 3.0~\AA, a marked increase in
magnitude with decreasing anion-cation separation is apparent. The fits to the other calculation levels 
have less fluctuation in the $4f$ crystal-field parameters.

The $5d$ crystal-field parameter $B^4(d)$ shows a smooth increase in
magnitude. This may be approximated by a power-law dependence of $R^{-5.5}_{\mathrm{Cl}}$. As would be expected, this dependence is
steeper than a simple point-charge crystal-field model \cite{NN89a}. Correspondingly, CsCaBr$_{3}$:Yb$^{2+}$ shows a smooth decrease
in magnitude of $5d$ crystal field parameter, due to the ligand configuration. This has an approximate dependence of $R^{-2.7}_{\mathrm{Cl}}$. 

We have already noted the discontinuity in $\Delta_E(fs)$ at 2.85~\AA\ for SrCl$_{2}$,
as the excited electron switches between $6s$ character at long anion-cation separations, and delocalized character at short separations. The $R^2(ds)$
and $R^3(ds)$ parameters are too uncertain to draw any
conclusions. However, the $G^3(fs)$ parameter does appear to exhibit a
discontinuity, dropping in value by a factor of 3 at the
discontinuity. If the excited electron becomes delocalized, as in the case of ITEs, it would be expected that the magnitude of the Coulombic interaction between
the excited electron and the $4f^{13}$ core would decrease, which indicates that this occurrence is a good description of excitonic behavior. 
The $R^2(ds)$ and $R^3(ds)$ parameters are more stable for the CsCaBr$_{3}$ system,
 but this may be due to the limited range of ion separations spanned by those calculations.

\subsection{Possible extensions}

Since the fits obtained are not exact, we investigated a number of 
possible extensions, such as considering any electron correlation effects
that could modify the crystal field, or allowing the $4f$ crystal-field parameters for the $4f^{13}5d$ and
$4f^{13}6s$ configurations to vary independently. None of these had a significant impact on
the fits.

\subsection{Comparison of fits}

The fits to experimental data, minima of the calculated curves for SrCl$_{2}$ and
energies at particular anion-cation separations ($> 2.7$ \AA) give comparable deviations, with the particular ion separation fit always slightly lower. 
It is notable that the fit at 3.1164~\AA\  gave a significantly lower
deviation than the fit to the minima or the fit to experimental
energies. Both the minima and the fit to experimental energies are
intended to be the zero phonon line positions, which are, in principle,
determined from the absorption spectrum. However, it is clear from the
avoided crossings in Fig.~\ref{fig:EPlot} that the eigenstates will be
very different for different ion separations, which may be why the fit to
the minima is not as good as some of the constant separation fits.

\section{Conclusions}

A ``crystal-field'' effective Hamiltonian has been constructed to model
the energy levels of SrCl$_{2}$:Yb$^{2+}$, and CsCaBr$_{3}$:Yb$^{2+}$, extending upon a $4f^{14} +
4f^{13}5d$ effective Hamiltonian to incorporate model states with
$6s$ character. The parameters were optimised by fitting to energy
levels determined by \emph{ab initio} calculations of S\'anchez-Sanz et al., for SrCl$_{2}$:Yb$^{2+}$~\cite{SaSeBa10a} 
and CsCaBr$_{3}$:Yb$^{2+}$~\cite{SaSeBa09} respectively. A good approximation can be achieved to both the
minima of the energy curves, and most of the energies at set anion-cation separations at each level of calculation presented in these references. The accuracy of the 
fits improved considerably at the positions of local minima of the energy curves determined in the \emph{ab initio} calculations. 

The $4f^{13}5d$ effective Hamiltonian parameters are
comparable to those determined from measured energy levels of SrCl$_{2}$:Yb$^{2+}$, with good agreement for the
spin-orbit and $5d$ crystal-field parameters; and a reasonable fit for the
Coulomb parameters, particularly the $G^k$ exchange parameters. The $4f$
crystal field parameters determined are of similar magnitude to the
values determined by Pan et al.~\cite{PaDuTa08}, but differ in relative sign. The $5d$
crystal field parameter increases in magnitude under contraction of the
ion separation. For the SrCl$_{2}$ system, this can be approximated by a power-law dependence of $R^{-5.5}_{\mathrm{Cl}}$. 

The $6s$ and $A_{1u}$ ITE potential wells can both be described
by the constructed effective Hamiltonian, with a different $G^3$ exchange
parameter in each regime. This corresponds well with the excited electron being in a localised or delocalised state respectively. 

Most of the parameters determined here should be transferable to other divalent ions (such as Tm$^{2+}$) and other crystals, (such as CaF$_{2}$). Of particular interest are the excitonic states of CaF$_{2}$:Yb$^{2+}$ and SrF$_{2}$:Yb$^{2+}$\cite{MoCoPe89,ReSeWeBeMeSaDuRe11}, which will be the subjects of future study.

\section*{Acknowledgments}
This work was supported by the Marsden fund of the Royal Society of New
Zealand, Grant No.\ 09-UOC-080.
L. S. and Z. B. acknowledge grant MAT2011-24586 from Ministerio de Econom\'\i a y Competitividad, Spain.

\section*{References}


\end{document}